\documentclass[final,3p,times,twocolumn]{elsarticle} 

\usepackage{graphicx}      
\usepackage{amsfonts}
\usepackage{gensymb}
\usepackage{amsmath}
\usepackage{amssymb}
\usepackage{amsthm}
\usepackage{algorithm}
\usepackage{algorithmic}
\usepackage{mathtools}
\usepackage{svg}
\usepackage{multirow}
\usepackage[shortlabels]{enumitem}
\usepackage{soul} 
\usepackage{chemformula}
\usepackage{siunitx}
\usepackage[version=4]{mhchem}
\usepackage[makeroom]{cancel}
\usepackage{stfloats}
\usepackage{pifont}
\usepackage{xcolor,colortbl}
\newcommand{\xmark}{\ding{55}}%

\definecolor{Color1}{rgb}{0.7804, 0.7373, 0.7529}
\definecolor{Color2}{rgb}{1, 1, 1}
\definecolor{Color3}{rgb}{0.8902, 0.8706, 0.8745}
\definecolor{Color4}{rgb}{0.9412, 0.9294, 0.9333}

\usepackage[linktocpage=true, colorlinks=true, linkcolor=blue,citecolor=blue, pdftitle={Process control}, pdfauthor={Noel Hallemans}]{hyperref} 

\setstcolor{red}\setulcolor{red} 

\begin{document}

\begin{frontmatter}



\title{Opportunities for real-time process control of electrode properties in lithium-ion battery manufacturing}

\author[inst1,inst5]{Noël Hallemans}
\author[inst1]{Philipp Dechent}
\author[inst1,inst5]{David Howey}
\author[inst2]{Simon Clark}
\author[inst3,inst5]{Mona Faraji Niri}
\author[inst3,inst5]{James Marco}
\author[inst4,inst5]{\\Patrick S. Grant}
\author[inst1,inst5]{Stephen R. Duncan\corref{cor1}}

\cortext[cor1]{Corresponding author} 
\ead{stephen.duncan@eng.ox.ac.uk}

\affiliation[inst1]{
            organization={Department of Engineering Science, University of Oxford},
            city={Oxford},
            postcode={OX1 3PJ},
            country={UK}
            }
  \affiliation[inst2]{
            organization={Battery Technology, SINTEF Industry},
            city={Trondheim},
            postcode={7031},
            country={Norway}
            }
\affiliation[inst3]{
            organization={WMG, University of Warwick},
            city={Coventry},
            postcode={7AL CV4},
            country={UK}
            }
\affiliation[inst4]{
            organization={Department of Materials, University of Oxford},
            city={Oxford},
            postcode={OX1 3PH},
            country={UK}
            }
\affiliation[inst5]{
            organization={The Faraday Institution},
            addressline={Harwell Campus},
            city={Didcot},
            postcode={OX11 0RA},
            country={UK}
            }

\begin{abstract}
Lithium-ion batteries (LIBs) have an important role in the shift required to achieve a global net-zero carbon target of 2050. Electrode manufacture is amongst the most expensive steps of the LIB manufacturing process and, despite its apparent maturity, optimised manufacturing conditions are arrived at by largely trial and error. Currently, LIB manufacturing plants are controlled to follow the fixed ``recipe'' obtained by trial and error, which may nonetheless be suboptimal. Moreover, regulating the process as a whole to conform to the set conditions is not widespread. Inspired by control approaches used in other film and sheet processes, we discuss opportunities for implementing real-time process control of electrode-related products, which has the potential to reduce the electrode manufacturing cost, CO2 emissions, usage of resources by increases in process yield, and throughput. We highlight the challenges and significant opportunities of implementing real-time process control in LIB electrode production lines.
\end{abstract}


\end{frontmatter}

\section{Introduction}
\begin{figure*}[ht]
    \centering
    \includegraphics[width=\textwidth]{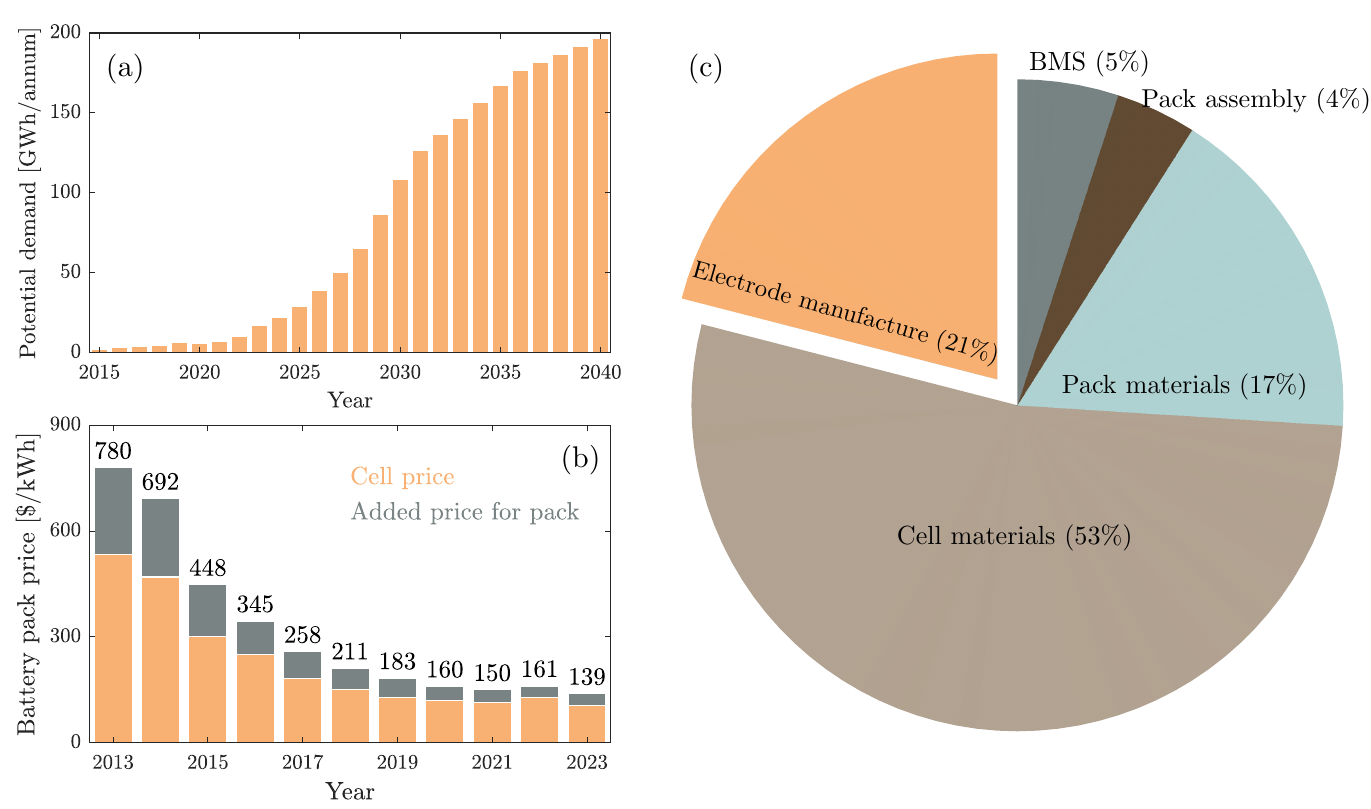}
    \caption{(a) Potential demand for UK-produced batteries \cite{FaradayReportSep2024}. Around 200 GWh (equivalent to 20 gigafactories each of 10 GWh) of supply will be required by 2040 in the UK to satisfy the need for batteries for electric vehicles and grid storage. (b) Volume-weighted average lithium-ion battery cell and pack price \cite{BloombergReportPrice139}. The survey includes 303 data points from commercial vehicles, passenger cars, buses, and stationary storage. Note that the rate of decrease in price has slowed, showing that batteries are becoming commodity products, i.e.\ price is increasingly related primarily to the volume produced. (c) Distribution of approximate battery pack cost \cite{UKbatteryStrategy}. Electrode manufacturing is the second most significant cost after cell materials.}
    \label{fig:batteryCostDemand}
\end{figure*}

Global demand for batteries is expected to increase dramatically over the next decade \cite{martins2021electric,fleischmann2023battery,BloombergReport} (see Fig.~\ref{fig:batteryCostDemand} (a) for the UK \cite{FaradayReportSep2024}). Primarily, this is driven by the need for energy storage in electric vehicles (EVs), which are strongly promoted by governments many of whom have set a net-zero carbon emissions target for 2050, together with the increasing need for flexible grid storage. Several authorities have set out their vision to achieve globally competitive domestic battery supply chains by 2030 \cite{WEF2019,USblueprint,EUbatteryActionPlan}. 

Lithium-ion batteries (LIBs) are attractive due to their high energy density and long cycle life, and despite challenges related to the supply of critical materials \cite{olivetti2017lithium}, the majority of battery production for EVs is still expected to be based on this technology in the critical time-frame to 2035. Although the cost of LIB-based cells has decreased significantly in recent years \cite{BloombergReportPrice139} (see Fig.~\ref{fig:batteryCostDemand} (b)), there is pressure to reduce costs further---for instance the UK ``Faraday Battery Challenge'' \cite{FaradayTargets} aims to further reduce battery pack cost by 2035. Given that in the short to medium term, the majority of battery production will be based on LIBs, to achieve this target it will be necessary to improve the throughput and yield of LIB electrode fabrication lines \cite{grant2022roadmap}, which are the second most expensive part of battery pack production \cite{UKbatteryStrategy,hawley2019electrode} (see Fig.~\ref{fig:batteryCostDemand} (c)). This can be achieved by limiting the amount of material that has to be scrapped because it does not satisfy the required quality specifications. Over the same time period, there is an aim to introduce a closed-loop recycling system so that 95\% of the material in a pack is recycled from end-of-life batteries \cite{FaradayTargets,harper2019recycling,ciez2019examining,sommerville2020review}, which will increase the variability of the feedstock re-entering electrode manufacturing lines. Significant amounts of energy are used in battery manufacturing, particularly on electrode lines, and there is a need to reduce the carbon footprint from the current level of up to the order of 100 kg CO$_2$-eq generated for each kWh of battery storage produced \cite{pucker2021greenhouse,kallitsis2024think}.

Real-time process control of electrode properties has the potential to address these issues. Although battery manufacturers are already making extensive use of feedback control on electrode lines, for example with speed control on drives and temperature control in ovens, they are currently applying closed-loop control to the \textit{process}, rather than to the \textit{product} (i.e., the electrodes themselves). This means that manufacturers are using open-loop control of the product by making electrodes to a standard ``recipe''. If suppliers of the raw materials can minimise variations in the feedstock, then---provided that the manufacturing process is tightly regulated---the required quality specifications can be achieved at acceptable yields by applying open-loop control. However, quality assessment of the product is typically carried out offline at the end of the process, so there is a delay before out-of-specification products are detected. Given that most manufacturing lines are over \SI{100}{m} long and line speeds exceed \SI{1}{m/s}, this can result in large amounts of scrapped material. To reduce the risk of producing too much scrap, processes tend to be operated conservatively, so line speeds are lower than necessary and excess drying is applied to ensure that the coating emerging from the ovens is dry, which wastes energy. There is a clear benefit in being able to measure product properties in real-time and use these measurements as part of feedback loops to make changes to the process in real time.

The current status of battery manufacturing is similar to the paper industry in the 1960’s, where the cost of paper was relatively high and manufacturers could sell all of the product that they could make. It was only in the 1980’s, when competition forced manufacturers to protect their margins by improving productivity, that feedback control started to be routinely applied in the industry \cite{dumont1986application,alma990150880730107026,stewart2003feedback,marshman2010energy}. There is currently insufficient battery manufacturing capacity with desired quality (compared with demand), so although there is pressure to improve the productivity of the process, it has not reached levels experienced by other commodity sectors. However, the driver for feedback control will come once batteries similarly become a ``commodity product'', where profit margins are reduced as the price of batteries decreases \cite{curry2017lithium}, while the costs of production, particularly energy, increase and the financial impact of scrapped material becomes significant. 

Apart from paper making, feedback control has also been successfully implemented in other sheet and film industries, including plastic film manufacturing \cite{hur2010modeling}, steel and aluminium sheet production \cite{Bryant1973,choi1994polynomial}, coating \cite{edwards1976coating}, and printing \cite{englund2008ink,seshadri2013modeling}. All of these industries use advanced machinery to manufacture commodity products with relatively small profit margins, and introducing feedback control has been necessary to maintain profitability by increasing production rates and reducing cost and start-up times while ensuring product quality \cite{featherstone2000identification}. 

Electrode manufacturing shows strong similarities to these film and sheet processes, however, closed-loop control of the product has not yet been extensively implemented in the production of LIBs, because it is likely to be more challenging than for film and sheet processes. Even if it cannot be implemented at present, the LIB industry could be future-proofing itself by ensuring that closed-loop control of the product can be incorporated into manufacturing lines at a later date. Closed-loop control will also allow production to be re-established quickly following start-up or disruption to minimise scrap material. This may be less of an issue for lines making single products, but is important for lines making multiple products. 

The aim of this paper is to raise awareness of the challenge of implementing closed-loop control of the product in electrode production lines. Moreover, we discuss opportunities for technology transfer and learning from other film and sheet processes.
\section{Electrode manufacturing}
\begin{figure*}
    \centering
    \includegraphics[width=0.95\textwidth]{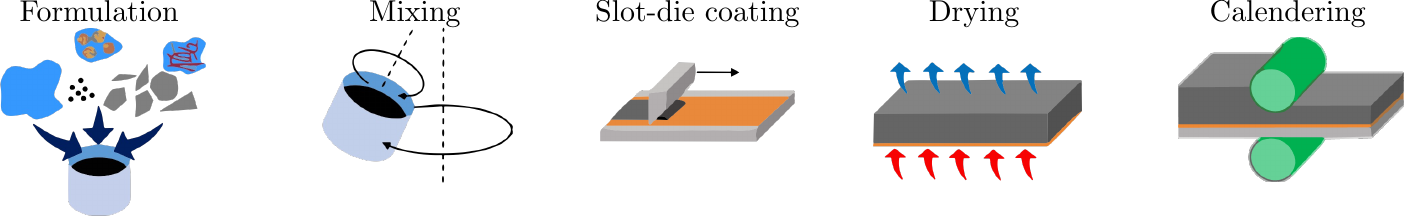}
    \caption{Illustration of the different steps in electrode manufacturing. Adapted from \cite{reynolds2021review}, CC-BY 10.1016/j.matdes.2021.109971.}
    \label{fig:electrodeManufacturingSteps}
\end{figure*}

Industrial electrode manufacturing lines use a roll-to-roll (R2R) approach in which a metallic foil current collector acts as a ``conveyor belt'' that may run at speeds up to \SI{100}{m/min}. A viscous slurry is formed by mixing the electrochemically active material, binder, and additives in a liquid, which is extruded through a slot-die onto the moving current collector. The slurry coating is then dried through an extended line of ovens, and calendered to the final thickness (30-\SI{120}{\micro m}) and porosity (30-40 vol\%) \cite{hawley2019electrode,li2021materials} (see Fig.~\ref{fig:electrodeManufacturingSteps} and relative costs in Table~\ref{tab:costManufacturingSteps}). The electrode is then slit, stacked, tabs applied, and packaged into the final cell. A ``recipe'' is followed, setting desired values for parameters such as mass fractions of powders, flow-rate of the coater, separate oven temperatures, calendering pressure, etc. The properties of the final electrode product---such as the energy capacity and the lifetime---are strongly dependent on this recipe \cite{gonccalves2022electrode}, which is often tuned based on trial-and-error or know-how. However, all aspects of the impact of the recipe on the electrode-related product properties and final cells are not yet fully understood \cite{reynolds2021review,duquesnoy2021machine,zhang2022applications,zhang2022review,reynolds2023impact,hidalgo2023design}. 
\begin{table}[]
    \centering
    \begin{tabular}{lcc}
 \rowcolor{Color1} &Cost & Energy consumption \\
 \rowcolor{Color2} Mixing & 28.2\%& 1.6\% \\
 \rowcolor{Color3} Coating & \cellcolor{Color4}  & 2.6\%\\
 \rowcolor{Color2} Drying & \cellcolor{Color4} \multirow{ -2}{*}{53.3\%}& 90.3\% \\ 
 \rowcolor{Color3} Calendering& 18.5\%& 5.5\% 
 \end{tabular}
    \caption{Relative cost and energy consumption of the different considered steps of electrode manufacturing \cite{liu2021current}. We note that drying has the highest cost and energy consumption.}
    \label{tab:costManufacturingSteps}
\end{table}
\subsection{Mixing}
Electrodes consist of a mixture of particulate active material (1-\SI{10}{\micro m}), conductive additive (usually carbon black), and polymeric binder (up to 5 vol\%) \cite{hawley2019electrode}. 
\begin{itemize}
    \item [-] Active material serves as a host for lithium-ions, and its mass fraction determines the gravimetric energy density of the battery.
    \item [-] Conductive additive ensures that electrons can percolate throughout the electrode and to/from the current collectors, and, hence, its mass fraction and distribution contribute to the internal resistance of the battery. 
    \item [-] Binder serves to improve the mechanical properties of the electrode---adhesion and cohesion---with an effect on the lifetime of the battery.
    \item [-] Residual porosity filled with electrolyte ensures lithium-ions can percolate throughout the electrode, and the total amount of porosity (and electrode thickness) relate strongly to the volumetric power density that can be achieved.
\end{itemize}
Optimising the fractions of these components requires a trade-off between energy capacity, power, internal resistance, and lifetime \cite{reynolds2023impact}. 

The slurry is formed by mixing the materials, which is often a multi-step process that may include dry mixing followed by one or two wet mixing steps in a solvent. The mixing aims to distribute the components uniformly by countering agglomeration and sedimentation, so that the slurry is stable and can be coated at high speed \cite{hawley2019electrode}. For graphite-based negative electrodes, water is used as the solvent and CMC and SBR (carboxymethyl cellulose and styrene-butadiene rubber respectively) as binders. The most commonly used solvent for LIB positive electrodes, which are typically based on lithium oxides or phosphates of different types, is N-methyl-2-pyrrolidone (NMP), which dissolves a PVDF (polyvinylidene fluoride) binder. However, NMP is hazardous, expensive, and is gradually being eliminated from usage \cite{hawley2020lithium}. Using water as solvent for positive electrodes too is a promising alternative for both environmental and financial reasons, although there are challenges for the more chemically active positive electrodes, such as lithium and transition metal dissolution, and associated corrosion of the current collector \cite{hawley2021enabling}. Finally, there are also approaches using no solvent at all (solvent-free processing or dry coating) in which polymeric binder is distributed using a warm-shearing process \cite{liu2023roll,matthews2024solvent,matthews2024impact}. Some of the negative electrodes of Tesla 4680 cylindrical cells, for instance, are made using solvent-free processing \cite{ank2023lithium}.

\subsection{Coating}
\begin{figure}
    \centering
    \includegraphics[width=0.45\textwidth]{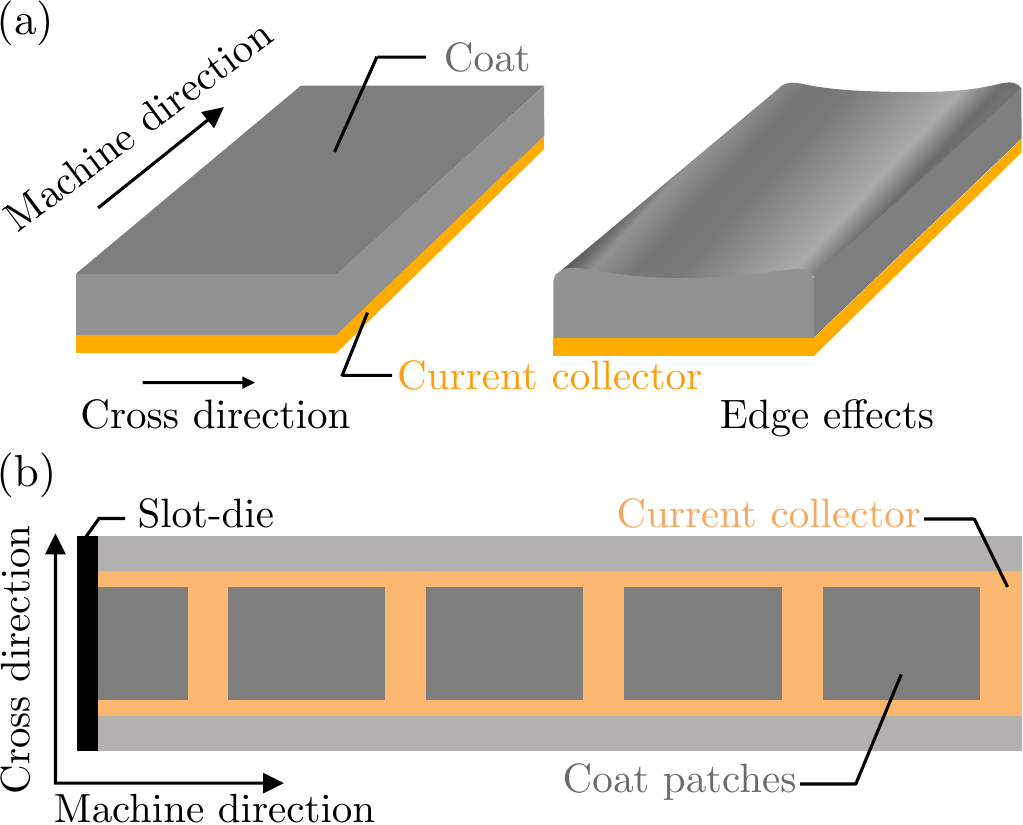}
    \caption{(a) Illustration of edge effects in electrode manufacturing, and indication of the machine and cross directions. (b) Top view of R2R current collector with slot-die coating patches of electrode product. The registration is the placement of the edges of these coat patches, both in machine and cross direction.}
    \label{fig:edgeEffectsAndPatches}
\end{figure}

The viscous slurry is coated onto the current collector (Cu for negative electrode, Al for positive electrode) through a slot-die or comma-bar. The pressure and width of the slot-die regulate the wet thickness of the electrode coating---typically in the range of 80-\SI{500}{\micro m}---and the coating is often applied to both sides of the current collector using multiple slot-dies. Coating widths may be tens of centimetres. To improve the production of electrode material, manufacturing lines must run at increasing speeds, but inevitably this makes it harder to ensure a uniform coating with target properties at all locations across the electrode \cite{schmitt2013slot}. For example, edge effects occur during electrode coating \cite{gong2024numerical}, whereby the cast thickness is not constant across the width (illustrated in Fig.~\ref{fig:edgeEffectsAndPatches}~(a)), and the gradient of these edge effects increases with the line speed \cite{schmitt2014slot}. Thickening of the edges leads to issues in subsequent manufacturing steps, such as cracking during calendering \cite{spiegel2022investigation}. Due to thicker edges, the critical ratio of the thickness of the negative and positive electrode (to ensure a energy-balanced cell) may not match the specification, which can lead to lithium-plating even at low currents \cite{deichmann2020investigating}, and hence, faster degradation, and safety issues \cite{attia2025challenges}. These edge effects may also cause problems at the downstream assembly steps. We will discuss below how this might be avoided using cross-directional control. Another important aspect is the geometrical registration of the coating---the position of the coating on the current collector---which becomes harder to control as line speeds are increased. 

For dry processing, extrusion can be used to coat dry particles and fibrillised or ``sticky'' binder onto the current collector \cite{bouguern2024engineering}. Single or twin screw extruders or rollers force the materials through a confined space, which also mixes them \cite{crawford2016recent}. In some cases, melt extrusion is applied, where materials are heated to a molten state that is forced through a mould that produces the electrode \cite{el2020melt}. Using a twin screw extruder for semi-dry and wet processing also provides advantages regarding design freedom, material and processing costs, easy upscaling, and continuous operation potential \cite{seeba2020extrusion,haarmann2021continuous,wiegmann2023semi}. 

\subsection{Drying}
\begin{figure}
    \centering
    \includegraphics[width=0.5\textwidth]{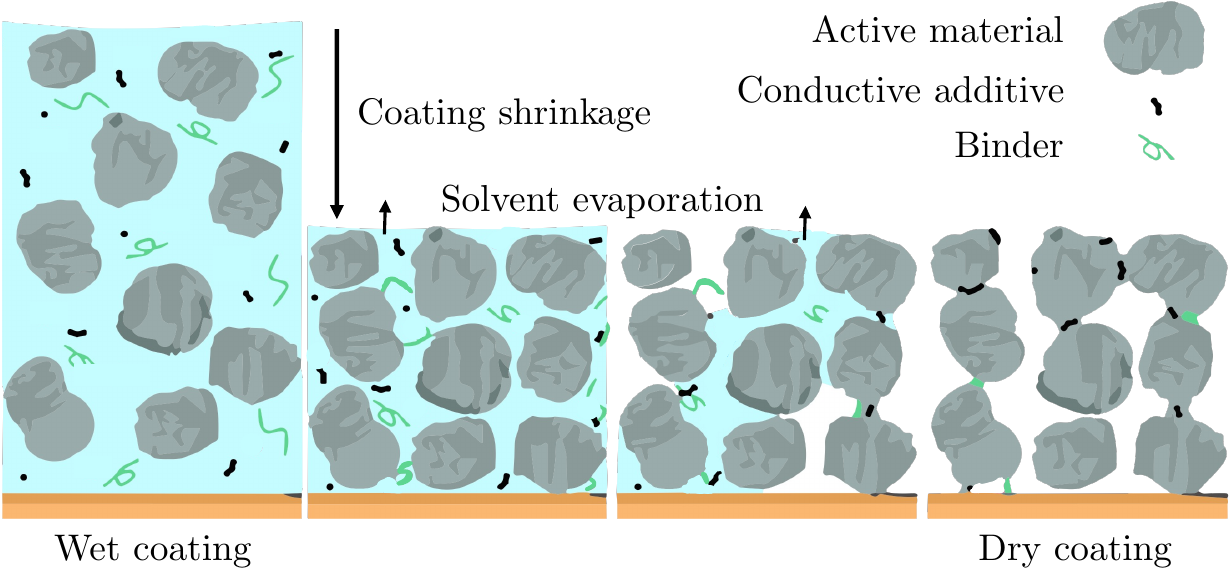}
    \caption{The electrode drying process. The conductive additive and binder are typically not distributed uniformly over the dry coating when the drying rate is too high. Illustration adapted from \cite{kumberg2019drying}.}
    \label{fig:drying}
\end{figure}

In solvent-based processing, the solvent is evaporated from the slurry (illustrated in Fig.~\ref{fig:drying}), and this is typically achieved by passing the coated electrode through a long series of ovens and applying conventional or infrared heating. Secondary drying is also sometimes performed after the calendering process \cite{zhang2022review}. Drying is the most energy-intensive step of electrode manufacturing (see Table~\ref{tab:costManufacturingSteps}) and when hazardous solvents, such as NMP, are used, these must be recovered. During drying, the solvent evaporates away from the current collector, and under poorly selected or regulated conditions, binder migration and/or sedimentation of particles may occur \cite{zhang2022review,jaiser2016investigation,kumberg2019drying}. The rate of drying is an important parameter for suppressing these effects and ensuring sufficient spatial uniformity of the different constituents and porosity in the final electrode. Slow drying leads to a more homogeneous spatial distribution, but limits throughput or requires longer lines, so a balance needs to be found between quality and throughput (including thickness). In practice, the slurry is often dried for longer than strictly necessary as a ``safety factor'' because real-time measures of ``dryness'' are unavailable.

\subsection{Calendering}
The volumetric energy density of the dry electrode is increased by densification, which is achieved by calendering, where heated hard steel rolls press the electrode coating. Densification also decreases the porosity, while increasing the tortuosity and enhancing electrical conductivity, and smoothing the free surface, all of which have effects on electrochemical and mechanical properties \cite{meyer2017characterization,gunther2020classification,ngandjong2021investigating}. A chosen line load can be applied by varying the roll gap between the rolls, the roller speed can be adjusted, and calendering can be performed at different temperatures. Once again, these parameters must be optimised to obtain the desired electrode properties.

\section{Control in electrode manufacturing}
Automating electrode manufacturing can be considered to be a control problem with the objectives of guaranteeing safety, maximising throughput, minimising cost, and ensuring product quality. To achieve these objectives, plants need to have:
\begin{itemize}[-]
    \item Actuators that can adjust process variables on the basis of signals provided by a controller;
    \item Sensors that can measure process variables.
\end{itemize}
The variables of interest for controlling the electrode manufacturing process are: 
\begin{itemize}[-]
    \item Product specifications: the properties the electrode must have. 
    \item Process settings (often referred to as set-points): the ``recipe'' that is expected to lead to the chosen product specifications. 
    \item Actuator signals: the control variables applied to the plant in order to follow the recipe. 
    \item Process parameters: the variables that should follow the process settings. 
    \item Product parameters: the properties of the electrode product that should satisfy the product specifications.
\end{itemize}
Examples of these variables are listed in Table~\ref{tab:SignalsControlLoop}.

\begin{table*}[]
    \centering
    \begin{tabular}{ll}
 \rowcolor{Color1} &Examples in electrode manufacturing\\
 \rowcolor{Color2} Product specifications &Electrode thickness, coat weight, porosity, specific capacity, binder distribution\\
\rowcolor{Color3} Process settings& Line speed, mixing time, rotational speed mixer, dryer temperatures, calendering pressure\\ 
\rowcolor{Color2} Actuator signals& R2R motor speed, coating slot width, current applied to infrared elements\\ 
\rowcolor{Color3} Process parameters &Line speed, mixing time, rotational speed mixer, dryer temperatures, calendering pressure\\ 
\rowcolor{Color2} Product parameters &Electrode thickness, coat weight, porosity, specific capacity, binder distribution
 \end{tabular}
    \caption{Signals of interest for the control of the electrode manufacturing process.}
    \label{tab:SignalsControlLoop}
\end{table*}

\subsection{Control of the process}
\begin{figure*}
    \centering
    \includegraphics[width=\textwidth]{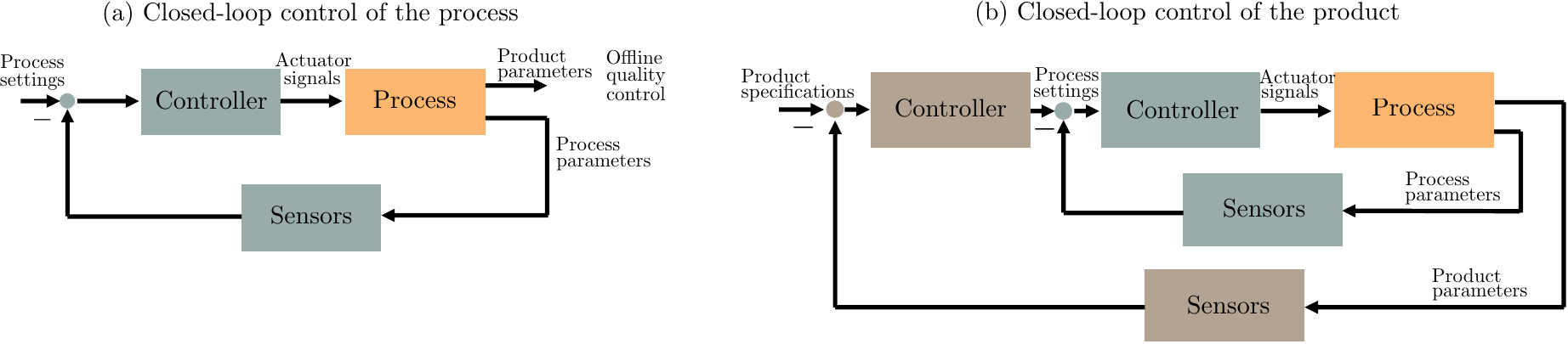}
    \caption{(a) Control loop for electrode manufacturing \emph{process}. Process settings (fixed recipe, e.g.\ line speed, dryer temperature) are inputs to the control loop; the goal is to have the actual process parameters follow these. To achieve this, sensor measurements of process parameters are fed back and actuators signals computed based on the difference between process settings and measured parameters. Quality control of the product is only performed offline. (b) Controlling the electrode manufacturing \emph{process} and \emph{product} in closed loop. Here also product parameters are fed back to adjust the process settings in real time.}
    \label{fig:feedbackLoops}
\end{figure*}

To date, only the electrode manufacturing \emph{process} is routinely regulated by closed-loop control. That is, the settings of the actuators are adjusted to follow the set-points of a fixed recipe for the product, typically through feedback, as illustrated in  Fig.~\ref{fig:feedbackLoops}~(a). The process settings (or recipe) leading to the desired product parameters are determined from know-how, trial-and-error, or using digital twins for the manufacturing process \cite{zanotto2022data}. The controller adjusts the actuator signals on the basis of the difference between the process settings and the measured process parameters, in order to ensure that the process parameters follow the desired process settings. Quality control of the electrodes is typically only performed offline, e.g. with electronic conductivity measurements on the finished electrodes.

The relation between the actuator signals and process parameters is a multivariable, dynamic, and nonlinear system. Often, this system is linearised about a set-point, such that a linear model can be used as a dynamic input-output relation. These linear models are then used, explicitly or implicitly, to design linear controllers for the process, so that the process parameters follow the process settings. The assumption of linearity is reasonable for a line that has been established at its desired operating conditions, where it is only necessary to regulate small deviations from steady-state values. However, as discussed below, the assumption of linearity may not be valid during large changes, for example at startup or during the transients associated with intermittent coating.

For reasons of simplicity, industry often uses proportional–integral–derivative (PID) controllers \cite{aastrom2001future}, generally without the derivative action. The gains of the terms in the controller are tuned, for instance, using the Ziegler-Nichols method or by minimising a chosen cost function \cite{aastrom2006advanced}. More complex control strategies can also be used, for instance optimising the actuator signal by minimising a cost function \cite{allwood2016closed}.

\subsection{Control of the product}
In battery manufacturing at present, the \emph{product}, unlike the process, is usually only controlled \emph{offline}. Some quality checks are performed online, for instance using cameras to monitor for cracks or visual inspections by operators, but most quality checks of the manufactured electrode are only performed once fabrication is finished. Hence, electrode defects can only belatedly be detected, leading to wasted materials, time, energy, and financial cost. Because of the need to increase production rates, not all electrode product may be subjected to quality control (although statistical quality control tools can be used \cite{montgomery2020introduction}), and some defects may not be detected at all. 

The current manufacturing status has the following drawbacks:
\begin{enumerate}[(i)]
\item It relies on a fixed recipe, often obtained through trial-and-error process design, which may not be fully optimised;
\item Manufacturing processes are usually run conservatively due to the risk of error;
\item Start-up times (e.g.\ transients in output properties and quality) can be long;
\item Production facilities are not particularly flexible when new types electrodes must be manufactured or when there is variability in the feedstock. 
\end{enumerate}
Hence, there is still scope for reducing electrode production cost, increasing production rates, and making production facilities more flexible. 

To control both \emph{process} and \emph{product}, a second (outer) control loop is required, where properties of the electrode product are measured or estimated \emph{during} the manufacturing process. The measurements are fed back to adjust the process settings (i.e.\ the recipe) in real time. Fig.~\ref{fig:feedbackLoops}~(b) shows an illustration of these two feedback loops, an inner one for controlling the \emph{process} and an outer one for controlling the \emph{product}. Linearised models at different set-points can be obtained for the inner (process) control loop \cite[Section 2.3]{featherstone2000identification} and, by feeding back product parameters, this system can in turn be controlled such that the product parameters follow the product specifications. The challenges here are the need for sensors and estimators that can provide information on the product, and for a controller design that adjusts the process settings based on this information. 

\section{Control of the product in related industries}
Although real-time process control of the product has not yet been extensively used for electrode manufacturing, it has been implemented successfully in other \emph{sheet and film} processes \cite{featherstone2000identification}. These include manufacture of steel and aluminium sheet \cite{Bryant1973,choi1994polynomial,grimble1995polynomial}, paper making \cite{dumont1986application,alma990150880730107026,stewart2003feedback,marshman2010energy}, polymer film extrusion \cite{wellstead1998identification}, offset printing \cite{englund2008ink,seshadri2013modeling}, and coating \cite{edwards1976coating}. The expensive machinery required for manufacturing these commodity products---up to \$1 million for a paper making machine, for instance---provided a strong economic incentive for optimising these processes. Implementing real-time process control of the product led to reduction of material consumption, larger production rates, shorter start-up time, improved product quality, and reduced energy consumption \cite{vanantwerp2007cross}. 

\begin{table*}
\centering
\begin{tabular}{lcccccc}   
\rowcolor{Color1}
&  \begin{tabular}{@{}c@{}}Paper \\ making\end{tabular} & \begin{tabular}{@{}c@{}}Plastic film \\ extrusion\end{tabular} & \begin{tabular}{@{}c@{}}Coating and \\ converting\end{tabular}& \begin{tabular}{@{}c@{}}Offset \\ printing\end{tabular} & \begin{tabular}{@{}c@{}}Steel \\ making\end{tabular} & \begin{tabular}{@{}c@{}}Electrode \\ manufacturing\end{tabular}\\  
\rowcolor{Color2} Typical cost [\$/ton] & 100-1000& 250 & 5000& 200-2000& 1000&20 000\\ 
\rowcolor{Color3}  Line speed [m/min]& Up to 2000& Up to 500& Up to 5& Up to 1000& 100& Up to 90\\ 
\rowcolor{Color2} Mixing& \checkmark&\checkmark& \checkmark& \xmark& \xmark&\checkmark\\ 
 \rowcolor{Color3} Slot-die coating& \xmark & \checkmark& \checkmark& \xmark& \checkmark& \checkmark\\
 \rowcolor{Color2} Drying/heat treatment& \checkmark& \checkmark& \checkmark& \checkmark& \checkmark& \checkmark\\
 \rowcolor{Color3} Calendering& \checkmark& \xmark& \xmark& \checkmark& \checkmark&\checkmark
\end{tabular}
\caption{Related industries and similarities with battery manufacturing. The costs per ton and line speeds are approximate.}
\label{tab:similarities}
\end{table*}
Electrode manufacturing is similar to sheet and film processing. These processes use a mechanism for conveying the sheet through the process at high speeds, with similar steps along the line. In Table \ref{tab:similarities}, we show that the steps of mixing, slot-die coating, extrusion, drying, and calendering in electrode manufacturing are also present in other film and sheet processes. 

\textit{Mixing} in electrode manufacturing is closest to mixing in paper making, which is relatively well understood and can already be controlled reasonably well. 

The \textit{slot-die coating} process in printing, coating, and converting is similar to the one used in electrode manufacturing. It is common in coating plastics to have multiple slot-dies for a two-sided coating, which is also necessary for electrode manufacturing. Offset printing does not mix materials, but rather coats the ink of each colour (cyan, magenta, yellow, and black) separately using an ink fountain and steel rolls \cite{englund2008ink}. The registration of each of these colours is usually monitored by printing small rectangles of these colours in a corner of the page. The temperature of the ink is also typically controlled to ensure the desired viscosity. 

\textit{Extrusion} is a common process in the polymer industry for manufacturing pipes, films, sheets, tubes, and rods \cite{lafleur2014polymer,abeykoon2014novelmodelbased}. The process is controlled by adjusting the pitch of the screw (which is a design choice), the screw speed, and the barrel temperature to obtain desired product properties \cite{tibbetts1998extrusion,jiang2012polymer,abeykoon2014novelrealtime,abeykoon2016single}. Typically a PID controller is used, but the process has a delay that depends on screw speed which complicates the design of the controller. Similar techniques could be used for the extrusion in dry-processed electrode manufacturing.

\textit{Drying} in electrode manufacturing is most similar to drying in the paper making industry \cite{stenstrom2020drying}, where water is removed from the product.

\textit{Calendering} is primarily used in paper making for smoothing the paper. However, in hot rolling of steel and aluminium sheet, a similar mechanism is used to change the thickness of the sheet by passing it between rollers, while in cold rolling, this affects the residual sheet tension. In both cases, this has a significant effect on microstructure, so, as calendering in electrode manufacture also has a large impact on the microstructure, it is more closely related to metal rolling.

The electrode manufacturing industry can transfer known technology and control strategies from film and sheet processes to implement real-time process control of the electrode product. This transfer has already started and knowledge from film and sheet processes is being considered for battery manufacturing. For example, Sony produced the first LIB electrodes with equipment used for manufacturing the plastic film for cassette-tapes, which had been standing idle \cite{stephan2021has}.

\section{Real-time process control of electrode product}
To control the electrode product, sensors are required to measure the product together with a model-based closed-loop control strategy to compute updated process settings for the actuators. 

\subsection{Sensors}\label{sec:Sensors}
Real-time process control requires sensors that can measure product-related quality metrics \emph{during} the manufacturing process, that is, sensors on the R2R system and in the ovens. Examples of such sensors include infrared spectroscopy to measure the composition of the slurry during drying (to track the solvent evaporation and avoid over-drying for example), ultrasound measurements to monitor the calendering process \cite{zhang2021situ,guk2024investigation}, digital imaging (e.g.\ charge coupled device cameras), lasers for thickness measurements, and radiation methods (e.g.\ X-ray and Beta) for weight measurements \cite{reynolds2021review}.

Various other metrology approaches, such as scanning electron microscopy, Raman spectroscopy, atomic force microscopy, and electrochemical testing have been used in literature to understand drying and calendering mechanisms \cite{zhang2022review,meyer2017characterization}, but these techniques are fragile and can realistically only be applied in off-site laboratory settings, so they are not suitable for application in industrial production of electrodes.

However, online sensors developed for monitoring sheet and film processes could be used in electrode manufacturing, including infrared, ultrasonic, beta-ray absorption, gamma-ray, X-ray absorption, X-ray fluorescence, microwave, visible light, magnetic, force distribution, and electric capacitance measurements \cite[Section 2.1]{featherstone2000identification}\cite{polyblank2014closed}. These sensors can measure thickness (caliper), moisture content, density (basis weight), opacity, and organic content of the coated product. Some sensors are expensive (up to \$300k for a single sensor \cite[Section 2.1]{featherstone2000identification}), so instead of having multiple individual sensors across the electrode, they can also be moved back and forth across the process. When this motion is combined with the movement of the current collector, the sensor traces a ``zig-zag'' pattern over the electrode \cite[Fig.5]{reynolds2021review}.

Online sensors can rarely measure directly within the bulk of the coated material and provide information about its microstructure \cite{allwood2016closed}. Therefore ``soft sensors'' are required to obtain information about properties that cannot be measured directly, e.g.\ porosity. Soft sensors are observers that use models to predict interior properties from those measured at the surface. To be reliable for closed-loop control, these soft sensors must be sufficiently accurate and a long estimation time may limit the control speed (see Section~\ref{Section:ReactionTime}).

\subsection{Actuators}
Actuators in electrode manufacturing include the rotation speed of blades for mixing, adjustment of the slot-die widths, motor speed of the R2R current collector, coating speed, applied current to the infrared heating elements at different locations in the ovens, the width of the gap between calendering rollers, the temperature at which calendering is performed, etc. These different actuators are positioned along the machine direction of manufacture, but can sometimes also be controlled differently along the cross-direction. However, actuators have limits in the speed of their dynamics, and hence, they cannot control rapidly varying disturbances.

\subsection{Model identification}
The algorithms within a controller are typically based on models that describe the response of the process and the product to changes in the actuator settings \cite{ayerbe2022digitalization}. Although these models will inevitably only be approximations to the actual response, the benefit of using feedback is that the controller corrects for model inaccuracies. Physics-based models for manufacturing processes are usually complicated, so simple linear low-order models, typically described by transfer functions \cite{pintelon1994parametric}, are often employed instead, for instance a first order transfer function depending on a gain and time constant. These can be estimated at different set-points through system identification, by applying a small but sufficiently exciting perturbation, measuring the response, and estimating the model parameters from the data \cite{ljungSystemidentification,pintelon2012system}. To maximise production, electrode manufacturing facilities have limited time to perform ``test'' experiments for modelling the plant, so extensive data gathering for model-free control \cite{coulson2019data} is unlikely to be feasible.

\subsection{Control strategies}
The process and product models are used to design feedback control that uses sensor measurements of the product to update actuator settings in real-time. As can be seen in Fig.~\ref{fig:feedbackLoops}, the input of the controller is an error signal, either the difference between process parameters and process settings for the inner loop, or the difference between product specifications and product parameters for the outer loop. The idea is to compute the actuator signal such as to make these differences small and minimising a cost function. In the most simple case PI(D) controllers can be used for both the inner and outer loops \cite{aastrom2004revisiting}. Since the electrode manufacturing process contains delays, controllers based on internal model control (IMC) \cite{garcia1982internal,morari1989robust} can be used, particularly Dahlin control \cite{tang2007study} and Smith predictors \cite{smith1957closer,brosilow1979structure,astrom1994new}, which can be considered a special case of IMC. More complex control strategies such as model predictive control \cite{kouvaritakis2016model} can also be used.

Different kinds of manufacturing plants may need different control strategies. For plants continuously running and producing one single product, the control should aim to reduce the effect of disturbances (e.g.\ changes in raw materials or plant temperature), which requires a regulator-type control strategy \cite{johnson1971accomodation}\cite[Section 2.3]{aastrom2021feedback}. For plants producing a range of electrodes (e.g.\ pilot plants), the aim is to establish the line quickly to minimise the amount of material that does not meet the required specification, which can be considered to be a tracking problem \cite[Section 2.4]{aastrom2021feedback}.

\subsection{Reaction time}\label{Section:ReactionTime}
Product control is inherently limited by constraints imposed by the actuator and sensor setup. The speed of response of the product control loop is constrained by the dynamics of the actuators and the sensors. Since electrode manufacturing occurs in a continuous roll process, the time taken to update the process settings corresponds to a distance of electrode product that cannot be controlled in real-time. The faster the data acquisition of the sensors (where the bottleneck would be the soft sensor estimation times) and the dynamics of the actuators, the shorter this distance. Cheaper sensor hardware with slower acquisition speeds will increase the distance of uncontrollable product. Typically, this means that product control feedback can only remove disturbances that are occurring over distances of the length of metres.

An additional problem is that product sensors are typically positioned down the line from the actuators, which corresponds to a delay between a disturbance or error being detected and corrective action being applied by an actuator. This also limits the speed of the control, which means that the outer control loop is typically slow compared to the inner loop that controls the process \cite{allwood2016closed}. 

\subsection{Cross-directional control}
Usually, it is necessary to achieve uniform product specifications not only in the direction of travel of the material through the process (the machine direction), but also in the cross direction, which is perpendicular to the movement of the electrode (see Fig.~\ref{fig:edgeEffectsAndPatches}). As discussed previously, a common cross-directional error is edge effects, where typically electrodes are thicker at the edges \cite{schmitt2014slot}. The pressure variation in the die that extrudes the material onto the current collector is carefully designed to ensure that the flow from the die is uniform across the sheet, but changes in feedstock and set up can cause cross-directional variations in flow, which can lead to errors at the edges (and elsewhere) of the electrode. To compensate for these effects, the width of the slot-die can be adjusted via a series of screws mounted across the die, which change the cross-directional profile of the die gap by bending the lip slice. Often, these screws are adjusted manually during start up and may not be subsequently adjusted during plant operation. However, automated adjustment of lip screws, either mechanically or via thermal expansion, are routinely used in paper making and plastic film industries \cite{vanantwerp2007cross}, and real time control of these screws is commonplace. To implement control of properties across the sheet, it is necessary to have a suitable sensor mechanism that can detect cross-directional variations. As described in Section~\ref{sec:Sensors}, this can either be an array of sensors positioned across the sheet or, more commonly, a single sensor that is mounted on a frame that allows the sensor to scan back and forth in the cross direction. 

In steel rolling, the width of the gap between rollers is adjusted in real-time via jacks to change the residual tension in cold rolling and sheet profile in hot rolling. It is possible that similar approaches could be used in the calendering process for electrode manufacturing \cite{mayr2022line}. However, some steel rolling mills change the roll gap by using fluid sprays arranged across the roll to change the thermal expansion of the rolls, which would not be appropriate for electrode manufacturing as the fluid would affect the coating. 

\subsection{Registration}
The registration of the coating refers to the location of the active material on the substrate---this is important in electrode manufacturing primarily because of the effect that misaligned coating can have on subsequent processing and assembly. Registration control is commonplace in processes such as printing and converting \cite{seshadri2013modeling,lee2015register,chen2018modeling}, where incorrect alignment affects the colour balance when combining colours from separate inks. The alignment is usually monitored using optical or camera-based sensors detecting small patches or colour bars printed on the edge of the product. Some electrode manufacturing lines use cameras to detect the location of the active material, although it appears that if a misalignment is detected, only manual adjustments are made to the flow through the slot-die. There is an opportunity for automatic adjustments to be made to flow to correct the registration.

\subsection{Intermittency of coating}
A key difference between electrode manufacturing for pouch cells and most other film and sheet processes is that the electrode material is not deposited continuously onto the current collector, but instead is deposited in distinct patches, with gaps between the blocks \cite{schmitt2014slot} (see Fig.~\ref{fig:edgeEffectsAndPatches}~(b)). In the coating and converting industries, this is sometimes referred to as patch coating or skip coating. 

Intermittent coating has a significant impact on the slot-die process, and also impacts drying and calendering, which leads to two problems. Firstly, it is important to have accurate registration of the patches in both the machine direction and the cross direction, otherwise once it is assembled, the length of the electrode in a cell may be incorrect, which directly affects the performance, energy capacity, and safety of the cell. A second problem is that there will be a transient at the start of each block, particularly in the flow of the coating material through the slot-die, which could affect the thickness of the coating in both the cross-direction and the machine direction. Although these variations can be controlled by adjusting the flow through the die or the width of the die gap, in practice, even if the variations can be sensed rapidly, the combined effect of the relatively slow response of the actuators coupled with the delay between the point of actuation and the location of the sensor means that it is unlikely that any errors can be corrected within a block of coating. Instead, this is a potential application for iterative or repetitive control, where the variations within a given block can be used to adjust the actuator settings for subsequent blocks. Iterative control has been used within steel rolling \cite{garimella94}, but is also widely used in a number of other applications, such as robotics undertaking repetitive tasks \cite{longman2000}, steel making \cite{RZEPNIEWSKI2008599} and chemical process industries \cite{wang2009}, where batch-to-batch or run-to-run control is important. 

A further issue is that many of the actuators associated with controlling the processes are inherently nonlinear. For most other film and sheet processes that are run continuously, once the line has been established at fixed operating conditions, the response of the actuators can be linearised around this operating point. This considerably simplifies the design of the feedback control system because it is based on a linear model of the process dynamics. This approach is unlikely to be applicable during the large transients at the start of a patch, and the length of the patch will not be sufficiently long to achieve a steady-state operation---meaning that the feedback control system will need to be designed for a nonlinear response. An additional problem for cross-directional control systems is that they are typically designed on the assumption that the spatial (cross-directional) response and the dynamic (machine direction) response can be decoupled \cite{vanantwerp2007cross}. This assumption is reasonable for controlling relatively small variations around a fixed operating point, but is unlikely to be valid during the transient at the start of coating a patch. Even if the response of the cross-directional operator is linear, if it is not reasonable to separate the spatial and dynamic components of the response, then the response is described by a large scale, multivariable system, which makes the controller design considerably more complicated. 

\section{Discussion}
Based on the current trend \cite{BloombergReportPrice139}, the price per kWh for batteries will continue to decrease, putting more pressure on improving the productivity of the manufacturing process by increasing throughput and reducing scrapped material that does not satisfy the quality specification. At the same time, increased variability of the feedstock associated with the use of recycled material makes achieving the required quality specification more challenging. The decision on whether to implement real-time feedback control will ultimately be based on commercial considerations, and in particular, whether the significant cost associated with implementing an online control system is justified by the improvements both in the profitability of the manufacturing line and the quality of the final product. For example, the high cost of drying relative to other steps (see Table~\ref{tab:costManufacturingSteps}) gives an economic incentive to start implementing real-time process control for this step and energy consumption could be reduced by stopping the drying process when sufficient solvent has evaporated, which can be tracked by sensors.

However, there are technical considerations that affect the decision to invest in online control. As this paper has highlighted, process control of product properties is routinely used in other film and sheet industries, but it has not yet been implemented extensively in battery electrode manufacturing. One reason may be that although the complexity of the electrode manufacturing process is similar to other industries such as paper making, which has comparable steps of mixing, extrusion, drying, and calendering, the performance of a battery is more dependent upon the microstructure of the electrode. Microstructure is tightly controlled in some other processes, such as steel rolling, while the arrangements of polymer chains is regulated in plastic film extrusion and fibre flocculation is controlled in paper making, but this is primarily to ensure the mechanical properties of the products, such as tensile strength and ductility. By contrast, electrodes are ``functional'' products and the microstructure has to ensure that the transfer of lithium-ions between negative and positive electrode occurs in a repeatable manner over many charge-discharge cycles, which is a more challenging requirement compared to ensuring the mechanical performance of the product.

The microstructure is also affected by \textit{interactions} between the different stages (mixing, coating, drying, and calendering), which are not yet fully understood, and make control complicated. Although simultaneous control of multiple actuators is successfully used in papermaking for regulating basis weight, moisture, and thickness, the interaction between properties is relatively well understood and the control loops for each property can be decoupled \cite{vanantwerp2007cross}. By comparison, the interactions between processes are more complex in electrode manufacturing.

Real-time control of microstructure is complex, primarily because it is difficult to sense microstructure online. For this reason, it is likely that online control for electrode lines will initially be applied to geometric properties, such as the thickness of the electrode or the location of the electrode material on the substrate (see Fig.~\ref{fig:edgeEffectsAndPatches}~(b)), or to the drying process. However, compared to other sheet processes, in electrode manufacture, there is a more complex interaction between control of these properties and the microstructure. For example, if the electrode has thick edges following extrusion, then the effect of calendering can result in non-uniform microstructure across the electrode. Although cross-directional control to regulate the thickness of the edges can have a direct benefit on the microstructure, the redistribution of material associated with the control may also change the microstructure at the edges, affecting the performance of the electrode. 

There is a practical problem associated with integrating the actuators and sensors required for online feedback control with the electrode manufacturing process. For example, this includes ensuring that there is space in the line to fit sensors, particularly sensors that scan across the line, and automating actuators such as lip slice screws. There is also a need to integrate any control system into existing supervisory control and data acquisition (SCADA) systems. This requires collaboration between the suppliers of the electrode lines (often each manufacturing step of the line is supplied by a different company), sensor manufacturers, control system suppliers, and the provider of the SCADA system. However, the challenges of system integration are not unique to electrode manufacturing and have already been successfully addressed by many industries, including steel rolling and papermaking. 

There is also a potential issue that the implementation of online control within battery manufacturing lines requires interdisciplinary expertise that covers both control engineering and electrode manufacturing. It is rare to find experienced personnel who can combine these two areas, so it will be necessary for control engineers to learn about electrode manufacturing, and vice-versa. Anecdotal evidence suggests that battery manufacturers are now addressing this by employing engineers who have experience of controlling other sheet and film processes.

\section{Conclusions}
The demand for battery production is increasing to meet customer demand and to achieve net-zero carbon targets typically in the 2040-2050 time frame. Electrode manufacturing is an expensive part of production and should be optimised to reduce cost and environmental impact in terms of energy consumption and use of resources. Currently, the electrode manufacturing process is controlled only to follow a fixed recipe, which may not be fully optimised. 

Related industries have optimised their manufacturing processes by implementing real-time process control of the product, by adjusting the fixed recipe in real-time based on sensor measurements of the product. This requires two feedback loops; an inner one to control the process and an outer one to control the product. Implementing this for film and sheet processes has led to increased production rates, reduced cost, and ensured quality of the product. Although the technology from other film and sheet processes has not yet been extensively applied to electrode manufacturing, there is an opportunity to reduce costs and improve quality in electrode manufacturing by transferring technology from these processes. Electrode manufacturers are including online sensing equipment in their manufacturing lines, but closing the loop to adjust the recipe in real-time has not yet been extensively implemented.

Implementing real-time process control of electrodes is challenging, and arguably more complicated than in other film and sheet processes. However, to be future-proof, electrode manufacturing lines should now build in the capability to implement real-time process control of the product. Even if this cannot be implemented right now, industry should consider how to incorporate this on their production lines. With this paper we aim to raise awareness of the technological challenge and opportunities related to implementing real-time process control of LIB electrode products.

\section*{Acknowledgements}
This research was supported by the Faraday Institution Nextrode (FIRG066) project, as well as the EU IntelLiGent (101069765) project through the UKRI Horizon Europe Guarantee (10038031). The authors would like to thank Alf Isaksson and Sreedhara Sarma for fruitful discussions. Any errors or wrong claims are due to the authors. For the purpose of Open Access, the authors have applied a CC BY public copyright licence to any Author Accepted Manuscript (AAM) version arising from this submission.


\bibliographystyle{elsarticle-num}
\bibliography{bibFile}

\begin{thebibliography}{100}
\expandafter\ifx\csname url\endcsname\relax
  \def\url#1{\texttt{#1}}\fi
\expandafter\ifx\csname urlprefix\endcsname\relax\def\urlprefix{URL }\fi
\expandafter\ifx\csname href\endcsname\relax
  \def\href#1#2{#2} \def\path#1{#1}\fi

\bibitem{martins2021electric}
L.~S. Martins, L.~F. Guimar{\~a}es, A.~B.~B. Junior, J.~A.~S. Ten{\'o}rio,
  D.~C.~R. Espinosa, Electric car battery: {An} overview on global demand,
  recycling and future approaches towards sustainability, Journal of
  environmental management 295 (2021) 113091.

\bibitem{fleischmann2023battery}
J.~Fleischmann, M.~Hanicke, E.~Horetsky, D.~Ibrahim, S.~Jautelat, M.~Linder,
  P.~Schaufuss, L.~Torscht, A.~van~de Rijt, Battery 2030: {Resilient},
  sustainable, and circular, McKinsey \& Company (2023) 2--18.

\bibitem{BloombergReport}
{Bloomberg}, Electric vehicle outlook 2023,
  \url{https://about.bnef.com/electric-vehicle-outlook/}, accessed: 24/11/2023
  (2023).

\bibitem{FaradayReportSep2024}
{The Faraday Institution}, {UK} gigafactory outlook ({September} 2024): {UK}
  electric vehicle and battery production potential to 2040,
  \url{https://www.faraday.ac.uk/wp-content/uploads/2024/09/Gigafactory-Report_2024_final_17Sept2024.pdf},
  accessed: 24/09/2024 (2024).

\bibitem{WEF2019}
{Global Battery Alliance}, A vision for a sustainable battery value chain in
  2030,
  \url{https://www3.weforum.org/docs/WEF_A_Vision_for_a_Sustainable_Battery_Value_Chain_in_2030_Report.pdf},
  accessed: 30/11/2023 (2019).

\bibitem{USblueprint}
{U.S. Department of Energy}, National blueprint for lithium batteries
  2021–2030,
  \url{https://www.energy.gov/sites/default/files/2021-06/FCAB%20National%20Blueprint%20Lithium%20Batteries%200621_0.pdf},
  accessed: 27/11/2023 (2021).

\bibitem{EUbatteryActionPlan}
{European Commission}, Strategic action plan on batteries,
  \url{https://eur-lex.europa.eu/resource.html?uri=cellar:0e8b694e-59b5-11e8-ab41-01aa75ed71a1.0003.02/DOC_3&format=PDF#:~:text=The%20Commission%20is%20promoting%20a,battery%20packs%2C%20and%20their%20use%2C},
  accessed: 29/11/2023 (2018).

\bibitem{BloombergReportPrice139}
{Bloomberg}, Lithium-ion battery pack prices hit record low of \$139/{kWh},
  \url{https://about.bnef.com/blog/lithium-ion-battery-pack-prices-hit-record-low-of-139-kwh/},
  accessed: 13/01/2025 (2023).

\bibitem{UKbatteryStrategy}
{UK Department for Business and Trade}, {UK} battery strategy,
  \url{https://www.gov.uk/government/publications/uk-battery-strategy},
  accessed: 27/11/2023 (2023).

\bibitem{olivetti2017lithium}
E.~A. Olivetti, G.~Ceder, G.~G. Gaustad, X.~Fu, Lithium-ion battery supply
  chain considerations: Analysis of potential bottlenecks in critical metals,
  Joule 1~(2) (2017) 229--243.

\bibitem{FaradayTargets}
{The Faraday Institution}, Powering {Britain}’s battery revolution: {Annual}
  report 2018/2019,
  \url{https://faraday.ac.uk/wp-content/uploads/2019/12/Faraday-Annual-Report-2018-19-part2-FINAL-1.pdf},
  accessed: 27/11/2023 (2019).

\bibitem{grant2022roadmap}
P.~S. Grant, D.~Greenwood, K.~Pardikar, R.~Smith, T.~Entwistle, L.~A.
  Middlemiss, G.~Murray, S.~A. Cussen, M.~J. Lain, M.~Capener, et~al., Roadmap
  on {Li}-ion battery manufacturing research, Journal of Physics: Energy 4~(4)
  (2022) 042006.

\bibitem{hawley2019electrode}
W.~B. Hawley, J.~Li, Electrode manufacturing for lithium-ion
  batteries—{Analysis} of current and next generation processing, Journal of
  Energy Storage 25 (2019) 100862.

\bibitem{harper2019recycling}
G.~Harper, R.~Sommerville, E.~Kendrick, L.~Driscoll, P.~Slater, R.~Stolkin,
  A.~Walton, P.~Christensen, O.~Heidrich, S.~Lambert, et~al., Recycling
  lithium-ion batteries from electric vehicles, nature 575~(7781) (2019)
  75--86.

\bibitem{ciez2019examining}
R.~E. Ciez, J.~Whitacre, Examining different recycling processes for
  lithium-ion batteries, Nature Sustainability 2~(2) (2019) 148--156.

\bibitem{sommerville2020review}
R.~Sommerville, J.~Shaw-Stewart, V.~Goodship, N.~Rowson, E.~Kendrick, A review
  of physical processes used in the safe recycling of lithium ion batteries,
  Sustainable Materials and Technologies 25 (2020) e00197.

\bibitem{pucker2021greenhouse}
J.~Pucker-Singer, C.~Aichberger, J.~Zupan{\v{c}}i{\v{c}}, C.~Neumann, D.~N.
  Bird, G.~Jungmeier, A.~Gubina, A.~Tuerk, Greenhouse gas emissions of
  stationary battery installations in two renewable energy projects,
  Sustainability 13~(11) (2021) 6330.

\bibitem{kallitsis2024think}
E.~Kallitsis, J.~J. Lindsay, M.~Chordia, B.~Wu, G.~J. Offer, J.~S. Edge, Think
  global act local: the dependency of global lithium-ion battery emissions on
  production location and material sources, Journal of Cleaner Production 449
  (2024) 141725.

\bibitem{dumont1986application}
G.~A. Dumont, Application of advanced control methods in the pulp and paper
  industry—{A} survey, Automatica 22~(2) (1986) 143--153.

\bibitem{alma990150880730107026}
A.~P. Featherstone, J.~G. VanAntwerp, R.~D. Braatz, Identification and control
  of sheet and film processes, Advances in industrial control, Springer,
  London, 2000.

\bibitem{stewart2003feedback}
G.~E. Stewart, D.~M. Gorinevsky, G.~A. Dumont, Feedback controller design for a
  spatially distributed system: The paper machine problem, IEEE Transactions on
  Control Systems Technology 11~(5) (2003) 612--628.

\bibitem{marshman2010energy}
D.~J. Marshman, T.~Chmelyk, M.~S. Sidhu, R.~B. Gopaluni, G.~A. Dumont, Energy
  optimization in a pulp and paper mill cogeneration facility, Applied Energy
  87~(11) (2010) 3514--3525.

\bibitem{curry2017lithium}
C.~Curry, Lithium-ion battery costs and market, Bloomberg New Energy Finance
  5~(4-6) (2017) 43.

\bibitem{hur2010modeling}
S.-H. Hur, R.~Katebi, A.~Taylor, Modeling and control of a plastic film
  manufacturing web process, IEEE Transactions on Industrial Informatics 7~(2)
  (2010) 171--178.

\bibitem{Bryant1973}
G.~F. Bryant, Automation of Tandem Mills, The Iron and Steel Institute,
  Amsterdam, 1973.

\bibitem{choi1994polynomial}
S.~G. Choi, M.~A. Johnson, M.~Grimble, Polynomial {LQG} control of back-up-roll
  eccentricity gauge variations in cold rolling mills, Automatica 30~(6) (1994)
  975--992.

\bibitem{edwards1976coating}
W.~Edwards, A.~Carlton, G.~Harvey, R.~Evans, P.~J. McKerrow, Coating mass
  control system design for a continuous galvanizing line, Automatica 12~(3)
  (1976) 225--235.

\bibitem{englund2008ink}
C.~Englund, A.~Verikas, Ink feed control in a web-fed offset printing press,
  The International Journal of Advanced Manufacturing Technology 39 (2008)
  919--930.

\bibitem{seshadri2013modeling}
A.~Seshadri, P.~R. Pagilla, J.~E. Lynch, Modeling print registration in
  roll-to-roll printing presses, Journal of Dynamic Systems, Measurement, and
  Control 135~(3) (2013) 031016.

\bibitem{featherstone2000identification}
A.~P. Featherstone, J.~G. VanAntwerp, R.~D. Braatz, Identification and control
  of sheet and film processes, Springer, 2000.

\bibitem{li2021materials}
J.~Li, J.~Fleetwood, W.~B. Hawley, W.~Kays, From materials to cell:
  State-of-the-art and prospective technologies for lithium-ion battery
  electrode processing, Chemical Reviews 122~(1) (2021) 903--956.

\bibitem{gonccalves2022electrode}
R.~Gon{\c{c}}alves, S.~Lanceros-M{\'e}ndez, C.~Costa, Electrode fabrication
  process and its influence in lithium-ion battery performance: State of the
  art and future trends, Electrochemistry Communications 135 (2022) 107210.

\bibitem{reynolds2021review}
C.~D. Reynolds, P.~R. Slater, S.~D. Hare, M.~J. Simmons, E.~Kendrick, A review
  of metrology in lithium-ion electrode coating processes, Materials \& Design
  209 (2021) 109971.

\bibitem{duquesnoy2021machine}
M.~Duquesnoy, I.~Boyano, L.~Ganborena, P.~Cereijo, E.~Ayerbe, A.~A. Franco,
  Machine learning-based assessment of the impact of the manufacturing process
  on battery electrode heterogeneity, Energy and AI 5 (2021) 100090.

\bibitem{zhang2022applications}
Y.~S. Zhang, J.~J. Bailey, Y.~Sun, A.~M. Boyce, W.~Dawson, C.~D. Reynolds,
  Z.~Zhang, X.~Lu, P.~Grant, E.~Kendrick, et~al., Applications of advanced
  metrology for understanding the effects of drying temperature in the
  lithium-ion battery electrode manufacturing process, Journal of Materials
  Chemistry A 10~(19) (2022) 10593--10603.

\bibitem{zhang2022review}
Y.~S. Zhang, N.~E. Courtier, Z.~Zhang, K.~Liu, J.~J. Bailey, A.~M. Boyce,
  G.~Richardson, P.~R. Shearing, E.~Kendrick, D.~J. Brett, A review of
  lithium-ion battery electrode drying: mechanisms and metrology, Advanced
  Energy Materials 12~(2) (2022) 2102233.

\bibitem{reynolds2023impact}
C.~Reynolds, M.~Faraji~Niri, M.~F. Hidalgo, R.~Heymer, L.~Roman, G.~Alsofi,
  H.~Khanom, B.~Pye, J.~Marco, E.~Kendrick, Impact of formulation and slurry
  properties on lithium-ion electrode manufacturing, Batteries \& Supercaps
  (2023) e202300396.

\bibitem{hidalgo2023design}
M.~Hidalgo, G.~Apachitei, D.~Dogaru, M.~Faraji-Niri, M.~Lain, M.~Copley,
  J.~Marco, Design of experiments for optimizing the calendering process in
  {Li}-ion battery manufacturing, Journal of Power Sources 573 (2023) 233091.

\bibitem{liu2021current}
Y.~Liu, R.~Zhang, J.~Wang, Y.~Wang, Current and future lithium-ion battery
  manufacturing, IScience 24~(4) (2021).

\bibitem{hawley2020lithium}
W.~B. Hawley, A.~Parejiya, Y.~Bai, H.~M. Meyer~III, D.~L. Wood~{III}, J.~Li,
  Lithium and transition metal dissolution due to aqueous processing in
  lithium-ion battery cathode active materials, Journal of Power Sources 466
  (2020) 228315.

\bibitem{hawley2021enabling}
W.~B. Hawley, H.~M. Meyer~III, J.~Li, Enabling aqueous processing for
  {LiNi$_{0.80}$Co$_{0.15}$Al$_{0.05}$O$_2$} ({NCA})-based lithium-ion battery
  cathodes using polyacrylic acid, Electrochimica Acta 380 (2021) 138203.

\bibitem{liu2023roll}
Y.~Liu, X.~Gong, C.~Podder, F.~Wang, Z.~Li, J.~Liu, J.~Fu, X.~Ma, P.~Vanaphuti,
  R.~Wang, et~al., Roll-to-roll solvent-free manufactured electrodes for
  fast-charging batteries, Joule 7~(5) (2023) 952--970.

\bibitem{matthews2024solvent}
G.~Matthews, S.~Wheeler, J.~Ram{\'\i}rez-Gonz{\'a}lez, P.~Grant, Solvent-free
  {NMC} electrodes for {Li}-ion batteries: unravelling the microstructure and
  formation of the {PTFE} nano-fibril network, Frontiers in Energy Research 11
  (2024) 1336344.

\bibitem{matthews2024impact}
G.~Matthews, B.~Meyer, C.~Doerrer, J.~Ramirez-Gonzalez, E.~Darnbrough,
  N.~Hallemans, D.~Armstrong, P.~S. Grant, Impact of binder content on particle
  fracture and microstructure of solvent-free electrodes for {Li}-ion
  batteries, Journal of Materials Chemistry A (2025).

\bibitem{ank2023lithium}
M.~Ank, A.~Sommer, K.~A. Gamra, J.~Sch{\"o}berl, M.~Leeb, J.~Schachtl,
  N.~Streidel, S.~Stock, M.~Schreiber, P.~Bilfinger, et~al., Lithium-ion cells
  in automotive applications: {Tesla} 4680 cylindrical cell teardown and
  characterization, Journal of The Electrochemical Society 170~(12) (2023)
  120536.

\bibitem{schmitt2013slot}
M.~Schmitt, M.~Baunach, L.~Wengeler, K.~Peters, P.~Junges, P.~Scharfer,
  W.~Schabel, Slot-die processing of lithium-ion battery electrodes—coating
  window characterization, Chemical Engineering and Processing: Process
  Intensification 68 (2013) 32--37.

\bibitem{gong2024numerical}
X.~Gong, J.~Han, F.~Yan, X.~Du, Numerical and experimental investigation on
  formation of the film for different die lip configurations in lithium-ion
  battery electrode slot-die coating, Journal of Coatings Technology and
  Research 21~(2) (2024) 481--492.

\bibitem{schmitt2014slot}
M.~Schmitt, P.~Scharfer, W.~Schabel, Slot die coating of lithium-ion battery
  electrodes: investigations on edge effect issues for stripe and pattern
  coatings, Journal of Coatings Technology and Research 11 (2014) 57--63.

\bibitem{spiegel2022investigation}
S.~Spiegel, T.~Heckmann, A.~Altvater, R.~Diehm, P.~Scharfer, W.~Schabel,
  Investigation of edge formation during the coating process of {Li}-ion
  battery electrodes, Journal of Coatings Technology and Research (2022) 1--10.

\bibitem{deichmann2020investigating}
E.~Deichmann, L.~Torres-Castro, J.~Lamb, M.~Karulkar, S.~Ivanov, C.~Grosso,
  L.~Gray, J.~Langendorf, F.~Garzon, Investigating the effects of lithium
  deposition on the abuse response of lithium-ion batteries, Journal of The
  Electrochemical Society 167~(9) (2020) 090552.

\bibitem{attia2025challenges}
P.~M. Attia, E.~Moch, P.~K. Herring, Challenges and opportunities for
  high-quality battery production at scale, Nature Communications 16~(1) (2025)
  611.

\bibitem{bouguern2024engineering}
M.~D. Bouguern, A.~K. Madikere Raghunatha~Reddy, X.~Li, S.~Deng, H.~Laryea,
  K.~Zaghib, Engineering dry electrode manufacturing for sustainable
  lithium-ion batteries, Batteries 10~(1) (2024) 39.

\bibitem{crawford2016recent}
D.~E. Crawford, J.~Casaban, Recent developments in mechanochemical materials
  synthesis by extrusion, Advanced Materials 28~(27) (2016) 5747--5754.

\bibitem{el2020melt}
S.~El~Khakani, N.~Verdier, D.~Lepage, A.~Pr{\'e}b{\'e}, D.~Aym{\'e}-Perrot,
  D.~Rochefort, M.~Doll{\'e}, Melt-processed electrode for lithium ion battery,
  Journal of Power Sources 454 (2020) 227884.

\bibitem{seeba2020extrusion}
J.~Seeba, S.~Reuber, C.~Heubner, A.~M{\"u}ller-K{\"o}hn, M.~Wolter,
  A.~Michaelis, Extrusion-based fabrication of electrodes for high-energy
  {Li}-ion batteries, Chemical Engineering Journal 402 (2020) 125551.

\bibitem{haarmann2021continuous}
M.~Haarmann, D.~Grie{\ss}l, A.~Kwade, Continuous processing of cathode slurry
  by extrusion for lithium-ion batteries, Energy Technology 9~(10) (2021)
  2100250.

\bibitem{wiegmann2023semi}
E.~Wiegmann, H.~Cavers, A.~Diener, A.~Kwade, Semi-dry extrusion-based
  processing for graphite anodes: Morphological insights and electrochemical
  performance, Energy Technology 11~(9) (2023) 2300341.

\bibitem{jaiser2016investigation}
S.~Jaiser, M.~M{\"u}ller, M.~Baunach, W.~Bauer, P.~Scharfer, W.~Schabel,
  Investigation of film solidification and binder migration during drying of
  {Li}-ion battery anodes, Journal of Power Sources 318 (2016) 210--219.

\bibitem{kumberg2019drying}
J.~Kumberg, M.~M{\"u}ller, R.~Diehm, S.~Spiegel, C.~Wachsmann, W.~Bauer,
  P.~Scharfer, W.~Schabel, Drying of lithium-ion battery anodes for use in
  high-energy cells: influence of electrode thickness on drying time, adhesion,
  and crack formation, Energy Technology 7~(11) (2019).

\bibitem{meyer2017characterization}
C.~Meyer, H.~Bockholt, W.~Haselrieder, A.~Kwade, Characterization of the
  calendering process for compaction of electrodes for lithium-ion batteries,
  Journal of Materials Processing Technology 249 (2017) 172--178.

\bibitem{gunther2020classification}
T.~G{\"u}nther, D.~Schreiner, A.~Metkar, C.~Meyer, A.~Kwade, G.~Reinhart,
  Classification of calendering-induced electrode defects and their influence
  on subsequent processes of lithium-ion battery production, Energy Technology
  8~(2) (2020) 1900026.

\bibitem{ngandjong2021investigating}
A.~C. Ngandjong, T.~Lombardo, E.~N. Primo, M.~Chouchane, A.~Shodiev,
  O.~Arcelus, A.~A. Franco, Investigating electrode calendering and its impact
  on electrochemical performance by means of a new discrete element method
  model: {Towards} a digital twin of {Li}-ion battery manufacturing, Journal of
  Power Sources 485 (2021) 229320.

\bibitem{zanotto2022data}
F.~M. Zanotto, D.~Z. Dominguez, E.~Ayerbe, I.~Boyano, C.~Burmeister,
  M.~Duquesnoy, M.~Eisentraeger, J.~F. Monta{\~n}o, A.~Gallo-Bueno, L.~Gold,
  et~al., Data specifications for battery manufacturing digitalization: current
  status, challenges, and opportunities, Batteries \& Supercaps 5~(9) (2022)
  e202200224.

\bibitem{aastrom2001future}
K.~J. {\AA}str{\"o}m, T.~H{\"a}gglund, The future of {PID} control, Control
  Engineering Practice 9~(11) (2001) 1163--1175.

\bibitem{aastrom2006advanced}
K.~J. {\AA}str{\"o}m, T.~H{\"a}gglund, Advanced {PID} control,
  ISA-International Society of Automation, 2006.

\bibitem{allwood2016closed}
J.~M. Allwood, S.~Duncan, J.~Cao, P.~Groche, G.~Hirt, B.~Kinsey, T.~Kuboki,
  M.~Liewald, A.~Sterzing, A.~E. Tekkaya, Closed-loop control of product
  properties in metal forming, CIRP {Annals} 65~(2) (2016) 573--596.

\bibitem{montgomery2020introduction}
D.~C. Montgomery, Introduction to statistical quality control, John Wiley \&
  Sons, 2020.

\bibitem{grimble1995polynomial}
M.~J. Grimble, Polynomial solution of the standard {H2} optimal control problem
  for machine control system applications: {LQG} gauge control problem design,
  Optimal Control Applications and Methods 16~(2) (1995) 77--104.

\bibitem{wellstead1998identification}
P.~Wellstead, W.~Heath, A.~Kjaer, Identification and control of web processes:
  {Polymer} film extrusion, Control Engineering Practice 6~(3) (1998) 321--331.

\bibitem{vanantwerp2007cross}
J.~G. VanAntwerp, A.~P. Featherstone, R.~D. Braatz, B.~A. Ogunnaike,
  Cross-directional control of sheet and film processes, Automatica 43~(2)
  (2007) 191--211.

\bibitem{lafleur2014polymer}
P.~G. Lafleur, B.~Vergnes, Polymer extrusion, John Wiley \& Sons, 2014.

\bibitem{abeykoon2014novelmodelbased}
C.~Abeykoon, A novel model-based controller for polymer extrusion, IEEE
  Transactions on Fuzzy Systems 22~(6) (2014) 1413--1430.

\bibitem{tibbetts1998extrusion}
B.~R. Tibbetts, J.~T.-Y. Wen, Extrusion process control: Modeling,
  identification, and optimization, IEEE Transactions on Control Systems
  technology 6~(2) (1998) 134--145.

\bibitem{jiang2012polymer}
Z.~Jiang, Y.~Yang, S.~Mo, K.~Yao, F.~Gao, Polymer extrusion: From control
  system design to product quality, Industrial \& engineering chemistry
  research 51~(45) (2012) 14759--14770.

\bibitem{abeykoon2014novelrealtime}
C.~Abeykoon, A novel soft sensor for real-time monitoring of the die melt
  temperature profile in polymer extrusion, IEEE Transactions on Industrial
  Electronics 61~(12) (2014) 7113--7123.

\bibitem{abeykoon2016single}
C.~Abeykoon, Single screw extrusion control: A comprehensive review and
  directions for improvements, Control Engineering Practice 51 (2016) 69--80.

\bibitem{stenstrom2020drying}
S.~Stenstr{\"o}m, Drying of paper: A review 2000--2018, Drying technology
  (2020).

\bibitem{stephan2021has}
A.~Stephan, L.~D. Anadon, V.~H. Hoffmann, How has external knowledge
  contributed to lithium-ion batteries for the energy transition?, Iscience
  24~(1) (2021).

\bibitem{zhang2021situ}
Y.~S. Zhang, A.~N. Pallipurath~Radhakrishnan, J.~B. Robinson, R.~E. Owen, T.~G.
  Tranter, E.~Kendrick, P.~R. Shearing, D.~J. Brett, In situ ultrasound
  acoustic measurement of the lithium-ion battery electrode drying process, ACS
  Applied Materials \& Interfaces 13~(30) (2021) 36605--36620.

\bibitem{guk2024investigation}
E.~Guk, M.~F. Niri, T.~A. Vincent, G.~Apachitei, C.~Briggs, B.~Gulsoy, S.~Chao,
  Z.~Guo, J.~E. Sansom, J.~Marco, Investigation of calendaring parameters on
  the microstructure of graphite anodes within lithium-ion batteries: Insights
  from ultrasonic testing, Journal of Power Sources 614 (2024) 235063.

\bibitem{polyblank2014closed}
J.~A. Polyblank, J.~M. Allwood, S.~R. Duncan, Closed-loop control of product
  properties in metal forming: A review and prospectus, Journal of Materials
  Processing Technology 214~(11) (2014) 2333--2348.

\bibitem{ayerbe2022digitalization}
E.~Ayerbe, M.~Berecibar, S.~Clark, A.~A. Franco, J.~Ruhland, Digitalization of
  battery manufacturing: current status, challenges, and opportunities,
  Advanced Energy Materials 12~(17) (2022) 2102696.

\bibitem{pintelon1994parametric}
R.~Pintelon, P.~Guillaume, Y.~Rolain, J.~Schoukens, H.~Van~Hamme, Parametric
  identification of transfer functions in the frequency domain---{A} survey,
  IEEE Transactions on Automatic Control 39~(11) (1994) 2245--2260.

\bibitem{ljungSystemidentification}
L.~Ljung, System Identification: Theory for the User, Prentice Hall, 1999.

\bibitem{pintelon2012system}
R.~Pintelon, J.~Schoukens, System Identification: A Frequency Domain Approach,
  John Wiley \& Sons, 2012.

\bibitem{coulson2019data}
J.~Coulson, J.~Lygeros, F.~D{\"o}rfler, Data-enabled predictive control: In the
  shallows of the {DeePC}, in: 2019 18th European Control Conference (ECC),
  IEEE, 2019, pp. 307--312.

\bibitem{aastrom2004revisiting}
K.~J. {\AA}str{\"o}m, T.~H{\"a}gglund, Revisiting the {Ziegler--Nichols} step
  response method for {PID} control, Journal of process control 14~(6) (2004)
  635--650.

\bibitem{garcia1982internal}
C.~E. Garcia, M.~Morari, Internal model control. {A} unifying review and some
  new results, Industrial \& Engineering Chemistry Process Design and
  Development 21~(2) (1982) 308--323.

\bibitem{morari1989robust}
M.~Morari, E.~Zafiriou, Robust Process Control, Prentice-Hall, 1989.

\bibitem{tang2007study}
W.~Tang, M.~Wang, Y.~Chao, L.~He, H.~Itoh, A study on the internal relationship
  among {Smith} predictor, {Dahlin} controller \& {PID}, in: 2007 IEEE
  International Conference on Automation and Logistics, IEEE, 2007, pp.
  3101--3106.

\bibitem{smith1957closer}
O.~J. Smith, Closer control of loops with dead time, Chemistry Engineering
  Progress~(5) (1957) 217--219.

\bibitem{brosilow1979structure}
C.~B. Brosilow, The structure and design of {Smith} predictors from the
  viewpoint of inferential control, in: Joint Automatic Control Conference,
  no.~16, 1979, p. 288.

\bibitem{astrom1994new}
K.~J. Astrom, C.~C. Hang, B.~Lim, A new {Smith} predictor for controlling a
  process with an integrator and long dead-time, IEEE Transactions on Automatic
  Control 39~(2) (1994) 343--345.

\bibitem{kouvaritakis2016model}
B.~Kouvaritakis, M.~Cannon, Model predictive control, Switzerland: Springer
  International Publishing 38 (2016) 13--56.

\bibitem{johnson1971accomodation}
C.~Johnson, Accomodation of external disturbances in linear regulator and
  servomechanism problems, IEEE Transactions on automatic control 16~(6) (1971)
  635--644.

\bibitem{aastrom2021feedback}
K.~J. {\AA}str{\"o}m, R.~Murray, Feedback systems: an introduction for
  scientists and engineers, Princeton University Press, 2021.

\bibitem{mayr2022line}
A.~Mayr, D.~Schreiner, B.~Stumper, R.~Daub, In-line sensor-based process
  control of the calendering process for lithium-ion batteries, Procedia CIRP
  107 (2022) 295--301.

\bibitem{lee2015register}
J.~Lee, J.~Seong, J.~Park, S.~Park, D.~Lee, K.-H. Shin, Register control
  algorithm for high resolution multilayer printing in the roll-to-roll
  process, Mechanical Systems and Signal Processing 60 (2015) 706--714.

\bibitem{chen2018modeling}
Z.~Chen, Y.~Zheng, T.~Zhang, D.~S.-H. Wong, Z.~Deng, Modeling and register
  control of the speed-up phase in roll-to-roll printing systems, IEEE
  Transactions on Automation Science and Engineering 16~(3) (2018) 1438--1449.

\bibitem{garimella94}
S.~Garimella, K.~Srinivasan, Application of repetitive control to eccentricity
  compensation in rolling, in: Proceedings of 1994 American Control Conference
  - ACC '94, Vol.~3, 1994, pp. 2904--2908 vol.3.

\bibitem{longman2000}
R.~W. Longman, Iterative learning control and repetitive control for
  engineering practice, International Journal of Control 73 (2000) 930–954.

\bibitem{RZEPNIEWSKI2008599}
A.~K. Rzepniewski, D.~E. Hardt, Development of general multivariable run-by-run
  control methods with application to a sheet metal forming process,
  International Journal of Machine Tools and Manufacture 48~(5) (2008)
  599--608, advances in Sheet Metal Forming Applications.

\bibitem{wang2009}
Y.~Wang, F.~Gao, F.~J. Doyle, Survey on iterative learning control, repetitive
  control, and run-to-run control, Journal of Process Control 19~(10) (2009)
  1589--1600.

\end{thebibliography}





\end{document}